# The temporal analogue of diffractive couplers

**Anastasiia Sheveleva, Pierre Colman, Christophe Finot**

*Laboratoire Interdisciplinaire Carnot de Bourgogne, UMR 6303 CNRS-Université de Bourgogne-Franche-Comté, 9 avenue Alain Savary, BP 47870, 21078 Dijon Cedex, France*

[*]<u>Corresponding author</u>:

E-mail address: anastasiia.sheveleva@u-bourgogne.fr

Tel.: +33 3 80395926

**Abstract:** Based on the space-time duality of light, we numerically demonstrate that temporal dispersion grating couplers can generate from a single pulse an array of replicas of equal amplitude. The phase-only profile of the temporal grating is optimized by a genetic algorithm that takes into account the optoelectronic bandwidth limitations of the setup.

## I. Introduction

The wave nature of light offers a very broad range of exciting possibilities to tailor its properties using simple transmissive elements. One well-known examples, which is found in any basics optics course textbook, is the use of a periodically structured component, namely a diffraction grating, to produce several spatial replicas of an incoming light beams [1, 2]. In the paraxial approximation, those copies, also called orders of diffraction, are equally spaced but present variations of intensities between replicas. That said, it is possible to achieve quasi-equal energy repartition between the spots of diffraction using more advanced design tools. This results into a very interesting components, known as a diffractive coupler, and that has now become widespread [3, 4].

If this concept of beam splitting has been fully developed in the field of spatial optics with for instance the kinoforms [5], it is also of interest to explore this idea in the temporal domain. Indeed, a very strong mathematical connection known as space/time duality exists between optical diffraction and the temporal evolution of light in presence of group velocity dispersion [6-8]. This concept has simulated both fundamental research regarding, for example, the Arago spot [9], two waves interferometers [10], and temporal Fresnel diffraction in fiber optics [11]. But also successful applications such as theory of temporal imaging [12], dispersive Fourier transform [13, 14], high-repetition rate sources, and related applications of the Talbot effect [15], and so many others.

In the present contribution, we first recall the basis of the analogy between the temporal and spatial aspects that constitute the framework for understanding the concept of dispersion grating [16] induced by simple temporal modulation schemes. Then, significant improvements can be achieved when the properties of the grating modulation are conveniently tailored. We compare for this purpose three different algorithms and show that the genetic algorithm appears the best-suited to achieve sequences ranging from three to nine identical and equally spaced pulses; or more complex sequences. Finally, we discuss the practical implementation of the concept proposed here, including the impact of the finite bandwidth of the optical phase modulator.

## II.    Principle of the approach

### a) Space-time analogy: concept of temporal gratings

Let us first start by recalling the basis of the space-time analogy and of the temporal gratings. Let consider the simple case where a beam having a 1D transverse field distribution of arbitrary shape $u_0(x)$ passes through the diffraction grating $G(x)$, eventually characterized by a spatial period $\Lambda$. The intensity distribution $I$ observed on a screen placed at distance $L$ is:

$$I(x,L) \propto \left| F.T.^{-1}\left( F.T.(u_0(x)G(x)) \exp\left(-i\frac{\lambda}{4\pi}Lk_x^2\right)\right)\right|^2, \qquad (1)$$

where $\lambda$ is a wavelength and $k_x$ is a transverse wavenumber. $F.T.$ and $F.T.^{-1}$ stand for the direct and reciprocal Fourier transforms, respectively. In the far-field approximation, Eq. (1) can be further simplified as [1] (note: the tilde stands for the F.T. of said variable) :

$$I(x,L) \propto \left| [u_0 * G]\left(\frac{2\pi}{\lambda L}x\right)\right|^2, \qquad (2)$$

Thus the spatial distribution of the intensity in the Fraunhoffer regime becomes a scaled replica of the spectrum of the input beam convolved by the grating. An example obtained for a sinusoidal phase modulation is provided in Fig. 1: the initial beam is split into an array of well-defined and equally spaced beamlets. In the most general case, $G(x)$ is a complex function which modifies both phase and amplitude of the input field. Depending on the profile of the diffraction grating, the output will therefore be a set of beams with an energy distribution that could be tailored. The main challenge in the design of diffractive couplers is therefore to find a periodic sequence leading to a spatial spectrum in line with the target. To reach this aim, different optimization processes were successfully proposed [17].

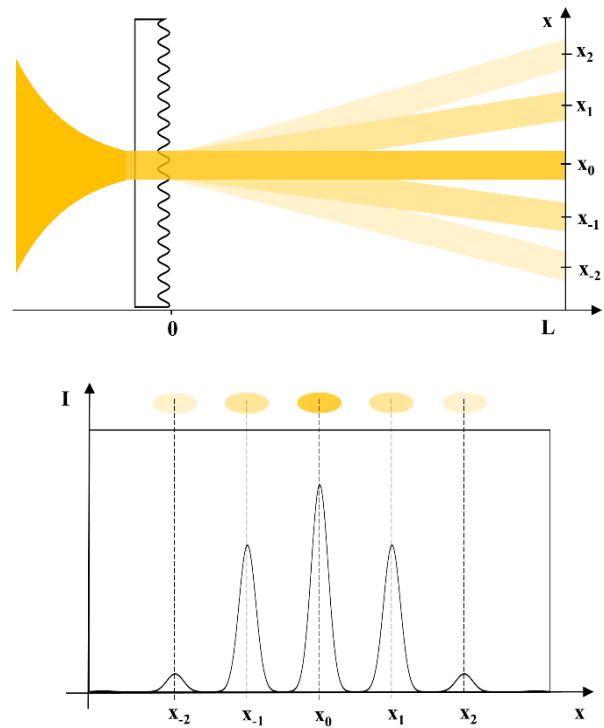

**Figure 1.** Schematic of the diffraction on a phase plate.

We would like to apply a similar concept in the temporal domain and find the pattern $G(t)$ of period $T_m$ that can split an incoming ultrashort pulse into a given number of replica with identical peak intensities. We therefore consider an input field $u_0(t)$ that propagates in a purely dispersive medium. Due to their ability to fully preserve the transverse spatial profile over very large distances, single-mode fiber components are very well suited. Moreover dispersion-compensating fibers and fiber Bragg gratings offer a large value of the second-order dispersion $\beta_2$. The longitudinal evolution of the temporal intensity profile can be predicted by [18] :

$$I(t,L) \propto \left| F.T.^{-1}\left( F.T.(u_0(t)G(t))\exp\left(i\frac{\beta_2}{2}L\omega^2\right)\right)\right|^2, \quad (3)$$

with $\omega$ the angular frequency. Equation (3) is therefore the temporal counterpart to Eq. (1). For short propagation distances, the temporal pattern leads to Talbot-carpet like pattern [15]. However, as the accumulated dispersion, hence propagation length, becomes more significant [19], well-separated temporal replica emerge from this dispersive grating [16]. Eventually the temporal intensity profile becomes a scaled copie of the optical spectrum, a process now well-known as the dispersive Fourier transform [13, 14] :

$$I(t,L) \propto \left|[u_0 * G]\left(\frac{-t}{\beta_2 L}\right)\right|^2 \quad (4)$$

This features have been experimentally validated in [16] where different transmission patterns typical of the telecom optical signal processing (Return-to-Zero modulation, Carrier-Suppressed Return-to-Zero modulation or sinusoidal phase modulation) have been tested, demonstrating the strong impact of the temporal pattern that is used to determine the dispersive grating.

### *b) Temporal gratings based on a sinusoidal phase modulation*

Despite ideal rectangular optical spectrum can be obtained using intensity modulators [20], we exclusively focus in the present contribution on dispersion grating induced by a continuous phase-only periodic modulation. In that context, the simplest phase profile that can be imprinted is of sinusoidal shape $\phi(t) = A_m \cos(\omega_m t)$, where $A_m$ is modulation depth and $\omega_m$ is the angular frequency of the modulation (the modulation period is defined as $T_m = 2\pi/\omega_m$). Consequently, the field after modulation by the phase grating is:

$$u(t) = u_0(t) \ G(t) = u_0(t) \exp(i A_m \cos(\omega_m t)), \quad (5)$$

which can be further rewritten by making a Jacobi-Anger expansion into [21, 22]:

$$u(t) = u_0(t) \sum_{l=-\infty}^{\infty} i^l J_l(A_m) \exp(i l \omega_m t) \quad (6)$$

where $J_l(A_m)$ are the Bessel functions of the first kind of order $l$ taken at the given value of the modulation depth. Such a sinusoidal phase modulation, followed by a subsequent propagation in a dispersive fiber has already been the key component of several ultrafast all-optical processing technics directly inspired by the time/space analogy. We can cite as examples the energy-efficient generation of train of ultrashort structures at very high-repetition rates formed from continuous wave input beam [23-25], the optical sampling based on the temporal analogue of the lenticular lenses [26], or various other regeneration schemes for optical transmissions [27, 28]. In contrast, our primary focus is the temporal intensity profile in the far-field when the modulation period is shorter than the duration of the pulse illuminating the grating. To obtain a solution for Eq. (4), one must Fourier transform Eq. (6), which will give the following optical spectra:

$$I(\omega) = |u(\omega)|^2 \propto \left| u_0(\omega) * \sum_{l=-\infty}^{\infty} i^l J_l(A_m) \delta(\omega - l \omega_m) \right|^2. \quad (7)$$

The optical spectrum is a convolution of the optical spectrum of the input field by a frequency comb spaced by $\omega_m$ and with amplitudes directly defined as square modulus of the Bessel function of the respective order. Equation (7) can be further simplified if we assume that the spectral extend of the optical spectra of the incoming pulse $|u_0(\omega)|^2$ is narrow compared to $\omega_m$, i.e. replica do not overlap spectrally:

$$I(\omega) \propto \sum_{l=-\infty}^{\infty} J_l^2(A_m) \left|u_0(\omega - l\,\omega_m)\right|^2. \tag{8}$$

As a consequence, the resulting temporal profile at a distance $L$ will be a made of a series of replica of the input pulse spectrum equally-spaced by the quantity $\tau_0 = |\beta_2 \omega_m L|$:

$$I(t,L) \propto \sum_{l=-\infty}^{\infty} J_l^2(A_m) \left|u_0\left(\frac{l\,\tau_0 - t}{\beta_2 L}\right)\right|^2. \tag{9}$$

Figure 2 illustrates the dispersive reshaping experienced by an input Gaussian pulse $u_0(t) = \sqrt{P_0}\exp\left(-t^2/2\sigma^2\right)$, where $P_0$ is the peak power and $\sigma = \tau/2\sqrt{\ln 2}$, $\tau$ being the full-width at half maximum (FWHM) - and equals here to $5\,T_m$ (Fig. 2(a1)). The dispersive grating is composed of a sinusoidal phase profile with a depth of $A_m = 1$ rad (Fig. 2(a1) top). Based on Eq. (3), we simulated the evolution of the pulse over a length of $3L_D$ with $L_D = T_m^2/|\beta_2|$ being a dispersive length. The fan-out behavior of the element can be observed in panel (a2): after an initial stage of propagation where Talbot-like carpet may exist, the different pulses emerge after a propagation of $L_D$, and have a good degree of separation after $2.5\,L_D$. Details of the intensity profile obtained after $3\,L_D$ are given in panel (a3): the output structure is composed of three peaks with amplitudes defined by $|J_l(A_m)|^2$.

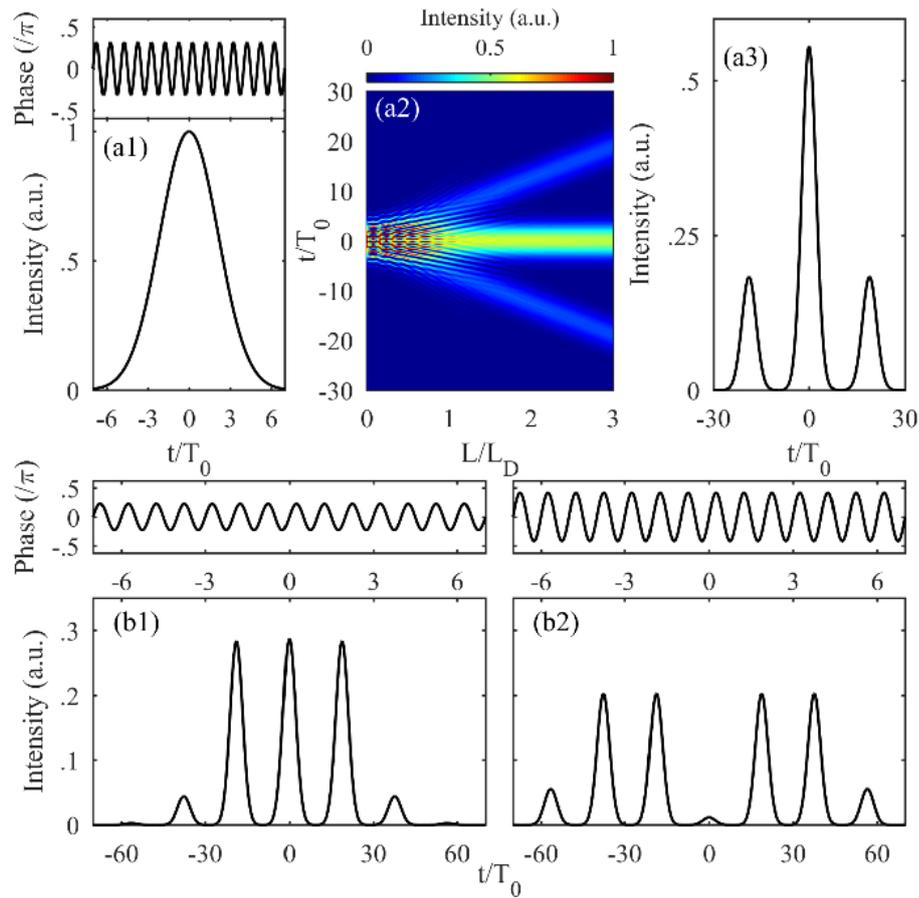

**Figure 2.** (a1) Input pulse with a width of 5 $T_m$ which is modulated with a sinusoidal phase profile at the modulation depth of $A_m = 1$ rad (displayed on top). (a2) Longitudinal evolution of the temporal intensity profile in the dispersive fiber. (a3) Output profile and the corresponding temporal phase. (b) Resulting waveforms from propagation in the same fiber of pulses modulated with sinusoidal phase at the modulation depths of 1.43 and 2.63 rad (panels 1 and 2, respectively)

A simple sinusoidal phase profile one can already result in some nicely tailored waveforms: a triplet (Fig. 2(b1), $A_m = 1.43$ rad) or a pair of pulse doublets (Fig. 2(b2), $A_m = 2.63$ rad). However, these waveforms are impaired by a non-negligible level of spurious sidelobes. For example, for $N = 3$ equalized peaks, the sidelobes contain 9.82% of the total energy and ±2 orders reach a peak power of 0.044 $P_0$. For the pair of doublets, 14.5% of the total energy in not comprised inside the main peaks.

## c) Optimization of the phase profile

If the simple sinusoidal phase modulation can already generate some nice temporal patterns, the possibility is however highly constrained by the values of the Bessel function of the first kind. More advanced periodic patterns *G(t)* are therefore required. A brute-force approach would be to imprint a large sinusoidal phase modulation $A_m$ in order to generate an electro-optic frequency comb [29, 30] that is further tailored using a programmable spectral shaper that adjusts line-by-line the amplitude of each spectral component [31]. Such an approach would however require the use of a phase-modulator with extremely large modulation capacity, which is rare and costly at high modulation frequency. Another solution implies a pair of phase modulators driven by two phase-shifted sinusoidal modulations [32]. Alternatives exist and, as it has been demonstrated in [17, 33, 34], a phase-only profile $G(t) = \exp(i\varphi(t))$ can be sufficient to tailor the output intensity profile and to obtain an array of pulses with equalized amplitudes. As it has been shown in the previous section the relative amplitude of the peaks is defined by a F.T. of the imprinted profile *G(t)*. Let us consider a phase-only mask of an arbitrary temporal shape, which F.T. reads:

$$\exp(i\varphi(t)) = \sum_{l=-\infty}^{\infty} a_l \exp(i\, l\, \omega_m t), \tag{10}$$

where $a_l$ are defined as:

$$a_l = \frac{1}{2\pi} \int_{-\pi}^{\pi} \exp(i\varphi(t))\exp(-i\, l\, \omega_m t)\, dt. \tag{11}$$

Therefore, we want to find the $\varphi(t)$ function which will give us a desired energy distribution among the $|a_l|^2$. In practice, we solve the problem backwards: first we search for the desired relation between the amplitudes $|a_l|^2$ by looking for coefficients $\mu_l$ and $\alpha_l$ that will lead to it [17]. Then, the phase profile can be restored using the formula:

$$\exp(i\varphi(t)) = \frac{\sum_{l=-n}^{n} \mu_l \exp(i\alpha_l)\exp(i\, l\, \omega_m\, t)}{\left|\sum_{l=-n}^{n} \mu_l \exp(i\alpha_l)\exp(i\, l\, \omega_m\, t)\right|}, \tag{12}$$

where $N = 2n + 1$ is the number of diffraction orders considered for the optimization.

The figure of merit attributed to a given solution directly depends on the optimization strategy which is used. The first target here is to generate *N*-peaks with equal amplitudes, distributed symmetrically on both sides of the input pulse, and that are lying on a zero background. Therefore, we have defined our fitness score function $\varepsilon$ as:

$$\varepsilon = \frac{\sum_{l=-n}^{n}\left||a_0|^2 - |a_l|^2\right| + \sum_{|l|>n}|a_l|^2}{\sum_{l=-\infty}^{\infty}|a_l|^2}, \qquad (13)$$

Here the score function is simply the normalized quadratic error with respect to the target (i.e. $a_0$ amplitude for the *N* central peaks, and zeros everywhere else). Another interesting problem could target a generation of a 'zero/one sequence' within *N*-diffraction orders, meaning that peaks with maximum and minimum amplitudes are alternated in the desired order. In this case the fitness score is defined as:

$$\varepsilon = \frac{\sum_{l \in P}\left||a_i|^2 - |a_j|^2\right| + \sum_{|l|>n}|a_l|^2 - \sum_{l \in M}\left||a_i|^2 - |a_j|^2\right|}{\sum_{l=-\infty}^{\infty}|a_l|^2}, \qquad (14)$$

where *P* and *M* are selections of 'ones' and 'zeros', respectively ($P, M \in [-n : n]$), and *i* is the order chosen to be the reference amplitude (hence $i \in P$, but *i* may differ from 0). In this variation of the score function, we decided to deviate from a pure metrics score (hence definite positive) by adding a third (negative) term in order to favor the realization of the zeros for $l \in M$ compared to the other background zeros |*l*|>*n*.

In order to solve this optimization problem in multidimensional space, we tested three different algorithms. Similarly to [33] a generalized reduced gradient algorithm has been used. This is the fastest method, but that suffers from poor results in case of complex score landscape, which exhibits in particular local minima. In this case the solution that is found strongly depends on the initial starting point. We also implemented a Nelder-Mead method [35] which is still fast and less prone to fall in local minima, but may output solutions that are not fully optimized solution (in brief the shrinkage procedure in the algorithm can make its final convergence very slow). In a sense, reduced gradient and Nelder-Mead algorithms are a bit complementary. Since the problem is defined by $2N$ parameters, it could be also important to investigate the largest proportion of the parameters space. Therefore, we used as the third optimization algorithm under test the genetic algorithm (GA) [36]. In our home-made GA each gene is composed of $\mu_l$ and $\alpha_l$ for $l = [-n : n]$. To increase the reproduction efficiency we have used three parents to produce each offspring [37]. Finally, to avoid condensation of the population into the same clone, we have been deliberately destroying a part of population in case genetic diversity decreases too much: this prevent all the individual to converge towards the same solution and hence force a wider exploration of the parameters space

## III. Examples of pulse sequences achieved by the temporal dispersive coupler

### a) 1 to 3 dispersive coupler

In order to verify the method discussed above, we first conducted numerical simulations to compute optimized phase profiles targeting the one to three coupling. The three tested algorithms converged to a same solution displayed in Fig. 3 that have a score of 0.079: the initial single pulse is efficiently converted into three identical Gaussian waveforms having identical peak powers. The optimum phase-only profile differs only slightly from the sinusoidal profile with a saturation affecting the extremum of the modulation. This small change is however sufficient to improve the result. Indeed, comparing results for sinusoidal (Fig. 2(b1)) and arbitrary (Fig. 3(c)) phase profiles we report a reduction of energy contained in the spurious replicas by 1.32 times and drop of amplitude of ±2 order by 2.44 times (the respective values for the tailored phase profile are 7.45% and 0.018 $P_0$). The efficiency of the setup, defined as the fraction of energy stored in the desired peaks over the whole space, is as high as 92.55 %. Results are here illustrated with an input Gaussian pulse, but the strength of the temporal coupler is that the ratio between the different the generated replicas does not depend on the input waveform. However, note that, except for the case of the Gaussian waveform for which the F.T. is also Gaussian, one may observe a change in the output pulse shape compared to the initial profile and some spurious low-amplitude ripple may for example appear in the case of Super-Gaussian temporal waveforms.

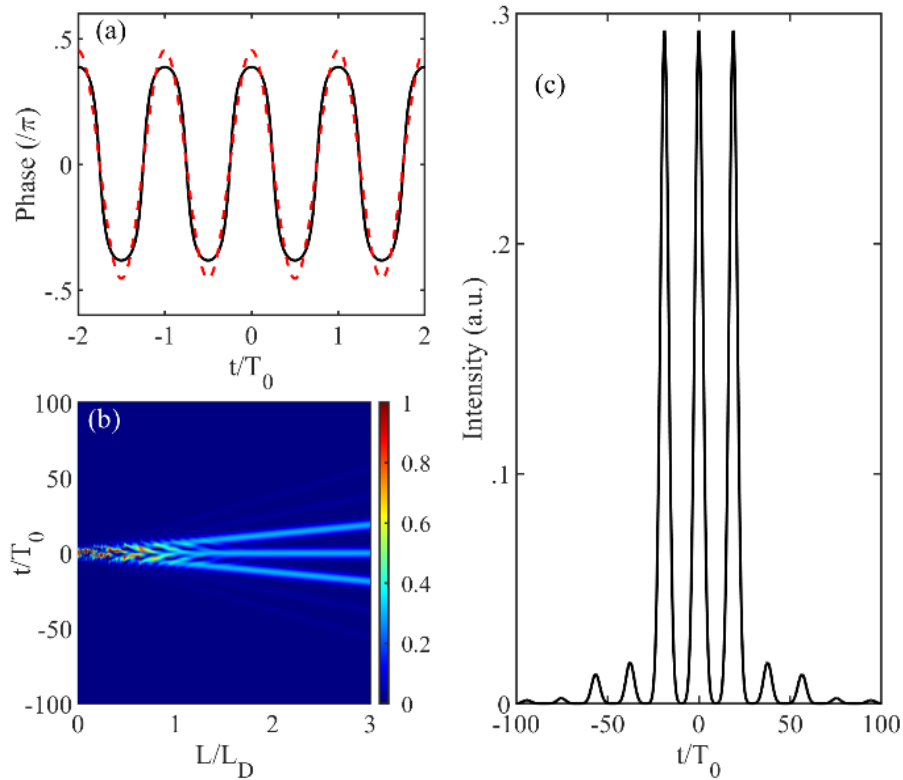

**Figure 3.** Results of numerical simulations targeting to achieve a set of 3. Panel (a) compares an optimized temporal phase profile (black line) to sinusoidal modulation with a depth of 1.43 rad (red dashed line). Panel (b) display evolution of the pulse in the dispersive optical fiber over 3 normalized propagation distances $L/L_D$. Panel (c) display output waveforms.

### *b) More advanced output sequences*

We then investigated the output sequences containing more output pulses (the properties of the input pulses and dispersive medium being kept identical). Examples of the results obtained for $N = 7$ pulses are reported in Fig. 4. We observe that the results obtained using either the genetic algorithm, the Nelder-Mead algorithm, or the least-square descent optimization differ. Indeed, when number of peak increases, the landscape of the score function has an increased complexity, the least-square methods may become trapped in one of the local minima. Due to its poorer convergence, the Nelder-Mead also failed to pinpoint the right position of the global minimum (So that even subsequent optimization using a gradient descent does not improve much the final result). As a consequence, the final solution is strongly influenced by the initial guess and the number of explored solutions is limited to a small fraction of the whole parameters space. On the contrary, the GA does not have such a limitation and therefore results in improved performance. It is able to find solutions that the Nelder-Mead and gradient descent may have missed. Indeed, the optimization score can be significantly reduced (0.033) compared to some local solutions obtained by the two other algorithms (0.24 and 0.07, for the Nelder-Mead and gradient descent, respectively). The resulting pulse sequence combining high uniformity of the peak powers and a low level of ghost pulses is fully satisfactory.

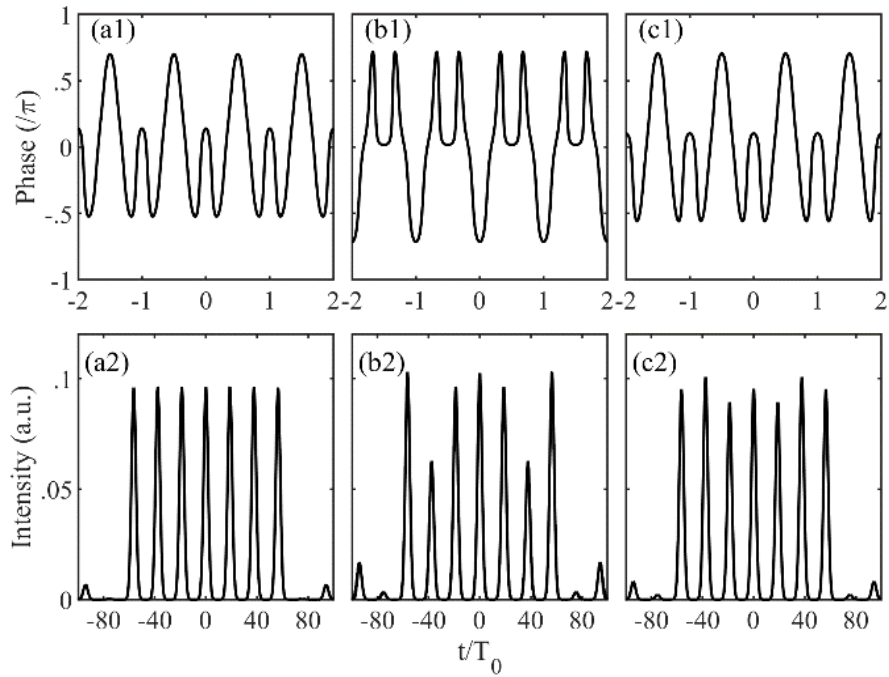

**Figure 4.** Comparison of optimized phase profiles made with (a) genetic algorithm, (b) Nelder-Mead algorithm, (c) least-square descent. Panels 1 and 2 display phase and output intensity profiles, respectively.

In the rest of this paper, due to its higher robustness, we used exclusively the genetic algorithm for our search of the best combination. Note however that if the GA gives out better results, the number of tested solutions (hence the number of generation times the population size) is much greater than for a single pass of either the Nelder-Mead or the gradient descent; so the question of the 'best' algorithm still remain open in the case where real-time operation is needed. Figure 5 summarize two other intensity profiles that can generated: a set of 9 peaks with equal amplitude and a set of 'zero/ones' with total number of 7 diffracted orders, where amplitudes of ±1 orders are minimized (1101011 sequence). The resulting phase profiles are depicted in panels 1 of Fig. 5. In both cases, longitudinal evolution of the intensity profile (based on Eq. (3), see panels 2) show that the temporal fan-out behavior of the element where the initial pulse reshapes upon linear propagation into a set of well-separated peaks. The output solutions are displayed in panels 3 of Fig. 5 and have the fitness scores of 0.059, -0.220, respectively. The last result has a negative score, meaning the remaining energy in the 'zeros' ($l \in M$) is much smaller than the natural level of the sidelobes ($|l|>n$). Both examples exhibit a high diffraction efficiency characterized by an uniform power distribution between the copies.

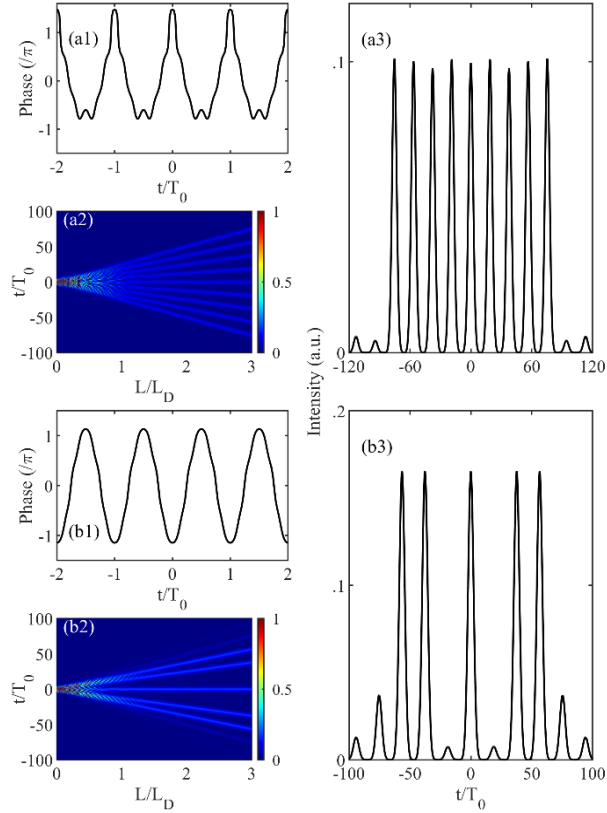

**Figure 5.** Results of numerical simulations targeting to achieve a set of 9 peaks with equal amplitudes and a 'zero/one' sequence with 7 peaks (panels (a) and (b), respectively). Panels 1 show a temporal phase profile imprinted on a Gaussian pulse of a width $5T_m$. Panels 2 display evolution of the pulse in the dispersive optical fiber over 3 normalized propagation distances. Panels 3 display output waveforms.

## IV. Influence of the bandwidth limitations of the output pulse sequence

In the spatial domain, a designer has to include a fabrication tolerancing analysis in his discussion [38]. In the temporal domain, the limitations are rather different. If a sinusoidal temporal phase modulation can eventually be synthetized at extremely high frequencies thanks to the cross-phase modulation induced by a beating of two optical wavelengths [39], using tailored shape of the phase profile is restricted in practice by the bandwidth of the optoelectronics components such as the phase modulators or the arbitrary waveform generators. In order to take realistic parameters into account, we consider here an initial modulation at a frequency $f_m = 20$ GHz. State-of-the-art modern equipment can currently handle frequency up to the 80 GHz. Therefore, it is important to know how such a limitation can influence the phase profile, hence the resulting waveforms. To study the influence of the BW-limitation, we took the solutions from the previous sections and then limited the RF spectra by a super-Gaussian profile of order 4 with a width of $4f_m$. Then, we considered propagation into a highly dispersive fiber element. It can be typically a normally dispersive fiber with a length $L$ of 58 km and a dispersion $\beta_2$ of 0.13 ps$^2$/m [40]; or a lumped element such as a commercially available chirped fiber-Bragg grating [41].

The results for the BW-limited profiles are presented in Fig. 6 (black lines). The resulting scores has increased up to 0.175, 0.598, –0.090 for $N$ = 3, 7 and 9 peaks (by factors 2.33, 2.44 and 10.14 in comparison to the ideal profiles, respectively). If the BW-limitation does not dramatically impair the first simplest two cases, we observed that the generation of a large number of equalized peaks is in turn much more impacted. Indeed the shape of the phase profile is more complex, so it becomes more BW-demanding.

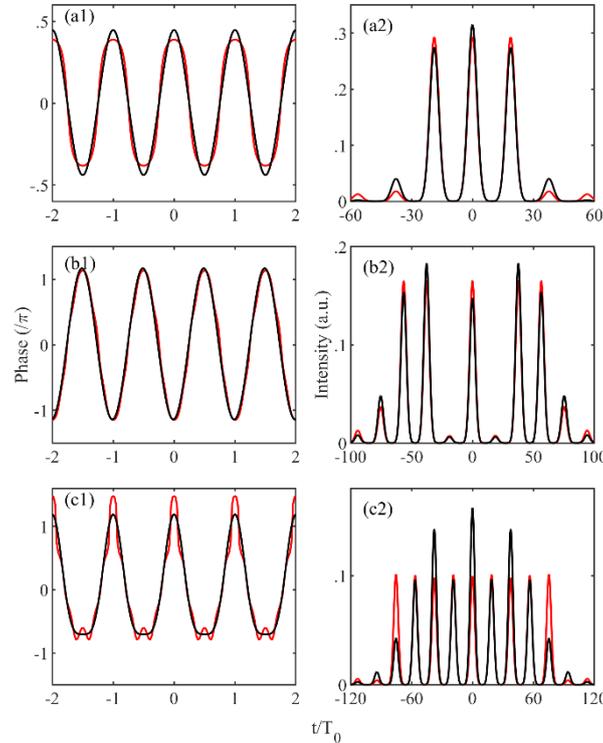

**Figure 6.** Comparison between the output waveforms modulated with ideal phase profiles and BW-limited ones (red and black lines, respectively) for $N$ = 3, 7 and 9 peaks (panels (a), (b) and (c), respectively). BW limitation assumed to be of a super-gaussian of order 4 with a width of $4f_m$.

However, because one of advantages of the GA is a flexibility in definition of the fitness score function, it is possible to include the BW-limitation directly into the simulation. Thus, we expect the simulation to adapt to realistic experimental constraints and look for solutions suitable. To demonstrate this approach, we have chosen to target a set of 7 peaks with equal amplitudes. First, we have obtained an ideal phase profile that delivered a score of 0.033 (refer to Fig. 4(a) that displays phase and amplitude profiles in panels 1 and 2, respectively). Then we have applied a BW-limitation of a super-Gaussian shape with a width of $4f_m$ and obtained a solution with a score of 0.966 (Fig. 7, red lines) which clearly fails the target of equalized intensity peaks. To restore the performance, we have included a BW-limitation into the GA and obtained an optimized solution which is depicted in black lines in Fig. 7. This optimized solution has achieved a score of 0.163 which reflects an improvement by 6.1 times in comparison to the ideal profile experiencing the BW-limitation.

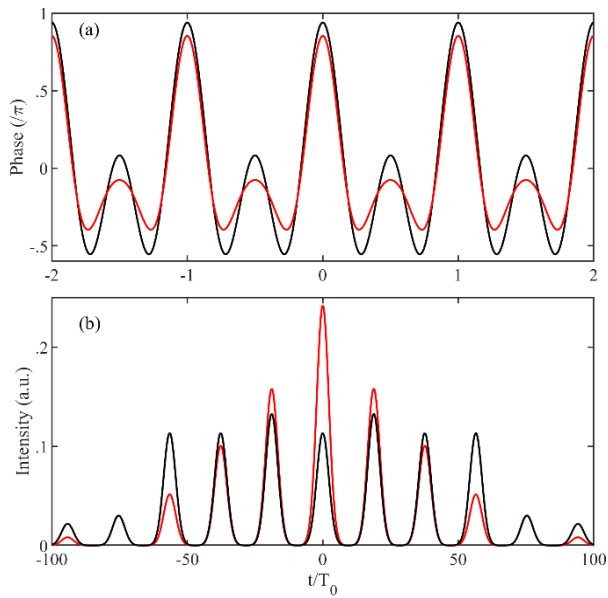

**Figure 7.** Comparison of phase profiles and the resulting waveforms (panels (a) and (b), respectively) of the BW-limited solution (red lines) to the BW-limited optimized solution (black lines). The limitation profile is a super-Gaussian of order 4 with a width of $4f_m$. To have a look on the ideal profile refer to Fig. 4(a).

## V. Conclusions

Based on the space-time duality of light, we have numerically demonstrated that a single optical pulse can be reshaped, through the transmission in a temporal phase-only grating, into a series of pulse of equal amplitude. The optimization of the phase-only grating is achieved thanks to a genetic algorithm that can also handle the detrimental optoelectronic limitations of the setup such as the finite bandwidth of the arbitrary waveform generator or the phase modulator. Similarly to the versatility brought by the spatial light modulators [34], the programmable arbitrary waveform generators, the temporal grating is fully reconfigurable and the sequence of pulses can be changed at will and the power ratio of the pulses in the sequence does not depend on the details of the incoming pulse. The system is also fully suitable for multiwavelength operation. Therefore, our proposed concept presents an interesting alternate to a set of classical beam splitting cubes and multiple delay-lines that could be very tricky to adjust in terms of delay and amplitude. It can be one of the building block of photonic setups targeting temporal super resolution [42] or temporal holography. One may also easily extend our concept to generate pulse sequence with different amplitudes that could be further used, through cross-phase modulation, to generate multi-focal imaging.

## Acknowledgements

The authors would like H. Rigneault for stimulating discussions regarding the space/time dual nature of light wave. This work was supported by the EUR EIPHI project (contracts ANR-17-EURE-0002), the ANR OPTIMAL project (ANR-20-CE30-0004), and the Région Bourgogne-Franche-Comté. C.F. was also supported by the Institut Universitaire de France. The numerical simulations relied on the HPC resources of DNUM CCUB (Centre de Calcul de l'Université de Bourgogne).